\documentclass[journal]{vgtc}                

\ifpdf
  \pdfoutput=1\relax                   
  \usepackage{graphicx}                
  \DeclareGraphicsExtensions{.pdf,.png,.jpg,.jpeg} 
\else
  \usepackage{graphicx}                
  \DeclareGraphicsExtensions{.eps}     
\fi%

\graphicspath{{figures/}{pictures/}{images/}{./}} 

\usepackage{microtype}                 
\PassOptionsToPackage{warn}{textcomp}  
\usepackage{textcomp}                  
\usepackage{mathptmx}                  
\usepackage{times}                     
\usepackage{cite}                      
\usepackage{tabu}                      
\usepackage{booktabs}                  

\usepackage{xcolor}

\usepackage{balance}

\newcommand{\name}{T-PickSeer}
\newcommand{\temporalView}{{temporal view}}
\newcommand{\temporalViewCapital}{{Temporal View}}

\newcommand{\mapView}{{map view}}
\newcommand{\mapViewCapital}{{Map View}}

\newcommand{\comparsionView}{{comparison view}}
\newcommand{\comparsionViewCapital}{{Comparison View}}

\newcommand{\rankView}{{rank view}}
\newcommand{\rankViewCapital}{{Rank View}}

\newcommand{\gsx}{\textcolor{black}}

\onlineid{1231}

\vgtccategory{Research}
\vgtcpapertype{application/design study}

\title{\name: Visual Analysis of \\Taxi Pick-up Point Selection Behavior}

\author{Shuxian Gu, Yemo Dai, Zezheng Feng, Yong Wang, and Haipeng Zeng}
\authorfooter{
\item
Shuxian Gu and Haipeng Zeng are with School of Intelligent Systems Engineering, Sun Yat-sen University. E-mail: gushx3@mail2.sysu.edu.cn; zenghp5@mail.sysu.edu.cn.
\item
 Yemo Dai is with Huawei Technologies Co., Ltd. E-mail: daiyemo@huawei.com.
\item
 Zezheng Feng is with Department of Computer Science and Engineering, The Hong Kong University of Science and Technology. E-mail: zfengak@connect.ust.hk.
 \item
 Yong Wang is with School of Computing and Information Systems, Singapore Management University. E-mail: yongwang@smu.edu.sg.
}

\shortauthortitle{Biv \MakeLowercase{\textit{et al.}}: Global Illumination for Fun and Profit}

\abstract{
Taxi drivers often take much time to navigate the streets to look for passengers, which leads to high vacancy rates and wasted resources. Empty taxi cruising remains a big concern for taxi companies. Analyzing the pick-up point selection behavior can solve this problem effectively, providing suggestions for taxi management and dispatch. Many studies have been devoted to analyzing and recommending hot-spot regions of pick-up points, which can make it easier for drivers to pick up passengers.
However, the selection of pick-up points is complex and affected by multiple factors, such as convenience and traffic management. Most existing approaches cannot produce satisfactory results in real-world applications because of the changing travel demands and the lack of interpretability. 
In this paper, we introduce a visual analytics system,~\name, for taxi company analysts to better explore and understand the pick-up point selection behavior of passengers.
We explore massive taxi GPS data and employ an overview-to-detail approach to enable effective analysis of pick-up point selection. 
Our system provides coordinated views to compare different regularities and characteristics in different regions.
Also, our system assists in identifying potential pick-up points and checking the performance of each pick-up point. Three case studies based on a real-world dataset and interviews with experts have demonstrated the effectiveness of our system.
} 


\CCScatlist{ 
 \CCScat{K.6.1}{Management of Computing and Information Systems}%
{Project and People Management}{Life Cycle};
 \CCScat{K.7.m}{The Computing Profession}{Miscellaneous}{Ethics}
}

\teaser{
  \centering
  \includegraphics[width=\linewidth]{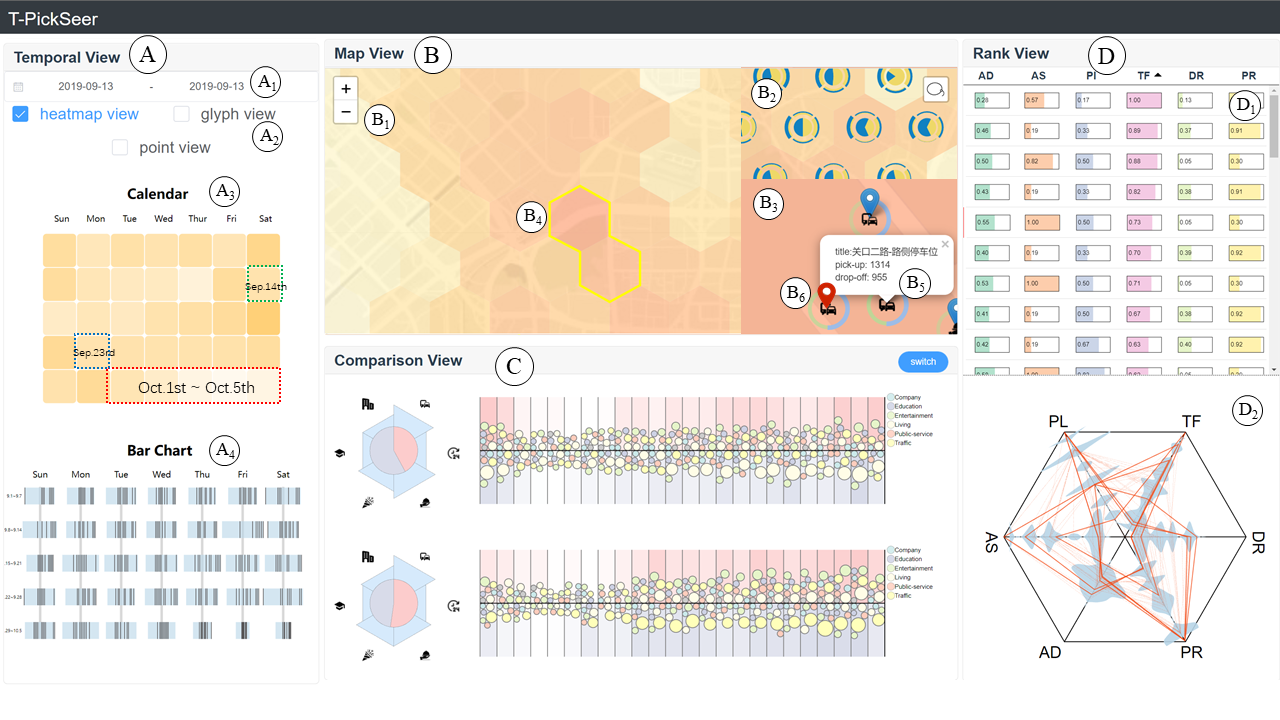}
  \caption{{\name} supports interactive exploration of pick-up point selection. (A) The temporal view comprises a time and view selection widget ($\rm A_1$, $\rm A_2$), a calendar heat map ($\rm A_3$), and a multiscale temporal chart ($\rm A_4$), enabling to configure the system and providing a temporal distribution of pick-up points. (B) The map view comprises a heatmap layer ($\rm B_1$), a glyph layer ($\rm B_2$), a POI layer ($\rm B_3$) and provides operations for selecting regions and points ($\rm B_4$, $\rm B_6$), facilitating multi-scale exploration. (C) The comparison view reveals patterns between different selected regions. (D) The rank view displays the performance of different POIs. The POIs are ranked based on a series of defined criteria on $\rm D_1$. $\rm D_2$ further summarizes the performance of selected POIs.}
	\label{fig:overview}

}

\vgtcinsertpkg

\begin{document}


\firstsection{Introduction}

\maketitle

With the rapid development of online taxi platforms (such as Uber and Didi),
taxi has become a popular travel choice~\cite{xiong2021understanding} and greatly improves the convenience of people's travel~\cite{wang2019passengers}. 
However, empty taxi cruising around the city causes a lot of energy waste every day, and taxi vacancy rate is high~\cite{mu2019recommend}. 
For example, taxi drivers often rely on personal experience to navigate the streets to look for passengers, which often takes much time though there are a large number of unfulfilled passenger orders in other areas~\cite{tang2017uncovering}.
To address this issue, some taxi companies offer online platforms for drivers to give advice on cruising. However, the information drivers can access is limited and has a lag, making the recommendation mechanism vague to them~\cite{wang2015taxirec}.
Taxi companies are in urgent need of more intelligent strategies to understand the needs of passengers, provide suggestions for drivers to cruise and improve the taxi service quality.

Analyzing the log data of pick-up point selection behavior of passengers can solve this problem effectively. 
First, understanding the spatio-temporal distribution of pick-up points helps taxi companies complete taxi scheduling. Through the discovery and comparative analysis of hot spots, taxi companies can issue dispatch instructions as soon as possible to effectively avoid drivers being trapped in congested traffic. 
Second, by analyzing the passenger travel~\gsx{characteristics} at different pick-up points, taxi companies can provide suggestions for drivers to better plan cruise routes.
Third, understanding the factors influencing passengers' pick-up point selection can help taxi companies adjust the settings of pick-up points. It can provide passengers with a comfortable ride experience.
Therefore, we focus on identifying potential patterns of pick-up point selection behavior of passengers hidden in a large number of GPS records.


Various research has been done to analyze passengers' taxi travel patterns.
Finding high-traffic areas in cities is a straightforward approach. Bi et al.~\cite{bi2021analysis} dug out the travel rules in hotspot regions. But the suggestions provided for cruising and dispatch are regional. More detailed information is lacking.
With the availability of large-scale GPS data, mining the GPS data helps taxi companies
gain more insights into the travel behaviors of crowds within cities. Ferreira et al.~\cite{ferreira2013visual} developed a visual analytics system to help explore large-scale GPS data collected from taxis. But it focused more on the mobility characteristics and big data query, rather than the pick-up selection. In addition, some studies have directly explored the cruising characteristics of high-income drivers. Gao et al.~\cite{gao2012visualization} analyzed the cruising routes of high-income drivers. Yuan et al.~\cite{yuan2020taxi} tried to find areas where high-revenue orders could be received.
But they cannot be applied to solve our problem. Different from routes or areas analysis, our analysis is multi-scale and focuses on pick-up point selection behavior.
Little research has been done on analyzing the pick-up point selection by passengers.

However, it is a challenging task to explore the pick-up point selection behavior of passengers due to three major reasons.
~\textbf{(1) Large-scale and multi-attribute spatio-temporal data.} The large scale of GPS data and the spatio-temporal attributes increase the difficulty of data analysis.
\textbf{(2) Dynamic urban transportation.} Urban transportation is dynamic and it is not easy to extract movement patterns from large-scale data, which makes it difficult to identify pick-up point preferences.
\textbf{(3) Multiple influencing factors.} Passengers' selection of pick-up point is influenced by many factors, such as travel purpose, traffic conditions~\cite{zhu2019pick}, convenience~\cite{7000580} and so on. It leads to complicated passenger preferences.
Due to these challenges, a fully automated analysis of pick-up point selection is difficult. Visual analysis, combined with both advanced computational power and human cognitive abilities, can be an effective solution for analyzing pick-up point selection behavior.

To address the above problems, we propose a visual analytics system to handle large volumes of time-varying traffic data, aiming at visual exploration of passengers' behavior for pick-up point selection in different regions. 
Analysts at taxi companies can use our system to obtain a series of useful discoveries, which can assist in the taxi operation management and dispatch guidance.
Specifically, we adopt a hierarchical exploration approach comprising city, region, and point scales. We combine GPS data with POI data to provide insight into pick-up point analysis. A novel comparison view is designed to facilitate the comparative analysis of different regions. In summary, the primary contributions of this paper are as follows:
\begin{itemize}
\item We propose an interactive visual analytics system for analysts in taxi companies to analyze passengers' pick-up point selection behavior at multiple scales (city, region, and point), featuring pattern exploration and potential pick-up points discovery.
\item A novel design is developed for easily comparing different pick-up point selection patterns. An augmented beeswarm graph is adopted to show numerous trips and corresponding POI information. Further, a stacked bar chart option is provided for better comparison of large-scale pick-up point selection data.

\item Three case studies using a real-world dataset, together with expert interviews, are conducted to evaluate the effectiveness of {\name} in empowering interactive exploration of passengers' pick-up point selection behavior.
\end{itemize}

\section{Related Work}
\label{sec:2}

This section reviews related research on
the analysis of pick-up selection~(Section~\ref{subsec: pick-up selection}) and visual analytics for traffic data~(Section~\ref{subsec: va of traffic data}).

\subsection{Analysis of Pick-up Selection}\label{subsec: pick-up selection}
The study of pick-up selection of passengers is beneficial to the management of taxi operation and future urban planning, which is of great significance to the urban structure, policy making, resource allocation, and so on. 
These studies can be mainly divided into two parts: pattern analysis and potential pick-up exploration.

The studies of pattern analysis are mainly committed to finding behavior patterns for pick-up selection and analyzing influencing factors. Traditional methods (e.g., statistics and regression) have been proved effective and many useful conclusions have been drawn ~\cite{ge2017urban}. But these suggestions are general and have trouble dealing with dynamic spatio-temporal data. More detailed information is desired. Then some research turns to the discovery and analysis of hotspots. 
Bi et al.~\cite{bi2021analysis} dug out the travel rules of different hotspots. However, further analysis of the differences between the hotspots is lacking.

Apart from these pattern analyses, some algorithms for finding and recommending potential pick-up locations are proposed. 
Various improved clustering algorithms are another commonly used technology. Berdeddouch et al.~\cite{berdeddouch2020recommender} utilized technologies of K-Means and regression to discover potential pick-up locations that are easier to find passengers. Zhang et al.~\cite{zhang2012recommending} improved spatio-temporal clustering based on K-Means and tried to find a set of personalized pick-up locations taking drivers' preferences into consideration by combining pick-up data and POI's attributions. Xu et al.~\cite{7000580} considered the information on taxi trajectory and improved the algorithm based on HotSpotScan and Preference Trajectory Scan algorithms. These methods take into account a variety of factors to recommend pick-up points. However, a deeper analysis of the importance of different factors is necessary.
Machine learning methods, such as deep learning~\cite{huang2019trec}, reinforcement learning~\cite{yang2020multiagent}, DeepFM~\cite{wang2021deepfm} also play an important role in potential location discovery. However, the model's output and access to information are limited, which confuses the user about the results. Although the above research has achieved good performance, these studies lack transparency.

In summary, existing studies are difficult for analysts who lack mathematical domain expertise to analyze. 
In addition, the current methods usually generate pick-up hotspots by various algorithms while further analysis of the differences between the hotspots is lacking, which also needs interpretation. 
It is difficult for them to intuitively display the temporal and spatial variation of the pick-up point.
In this paper, we focus on the visual analytics of the taxi pick-up data and try to find different patterns of pick-up selection, which help with the setting and recommendation of the pick-up point. We designed multi-scale analysis flow and interactive operations to analyze pick-up point selection, which can help users to conduct multidimensional analysis and judgment themselves.

\subsection{ Visual Analytics of Traffic Data}\label{subsec: va of traffic data}
Thanks to the development of fruitful advanced location-sensing technologies for collocating a vast amount of traffic data, the status of the moving objects, is recorded from both spatial and temporal dimensions~\cite{feng2022survey}.
The analyses of these data have made many contributions to solving urban problems, such as bus route planning~\cite{weng2020towards}, and taxi operation~\cite{ zong2018taxi, jiang2018impact}. 
However, due to the complexity of urban problems and the multidimensionality of traffic data~\cite{feng2020topology}, some of these methods may not perform well without the involvement of domain experts. 
The combination of data visualization and urban computing methods enables experts to explore traffic data interactively~\cite{deng2023survey}.

Data query focuses on developing new visual query models to quickly query traffic data and explore traffic information. Representative works include~\cite{al2016semantictraj,huang2019natural,ferreira2013visual,wang2014visual}. Filtering, sampling, aggregation, etc. combined with interaction~\cite{ chen2017vaud} can quickly present the results the user wants.
What's more, the main task that most studies focus on is pattern mining, committed to enabling analysts to obtain patterns and insights from big data~\cite{lu2016exploring}. 
\gsx{Zeng et al.~\cite{zeng2017visualizing} explored the relationship between human mobility and POIs. Deng et al.~\cite{deng2021visual} studied the cascades of spatial contagions.}
At last, different from pattern mining, some studies have tried to take advantage of the information extracted from pattern exploration for decision making. Weng et al.~\cite{weng2020towards} proposed a visual analytics system to generate optimal transit routes interactively. Liu et al.~\cite{liu2016smartadp} solved the problem of comparing solutions rapidly for billboard placements. Visualization helps to obtain useful information and interaction is allowed to participate in the process of generating decisions, making the system perform better than the algorithms. 

The success of visual analytics is inseparable from the help of three factors: the proper visual representation of data, good comparison design, and human interaction.
Firstly, various proper and novel visualization designs of spatio-temporal data have been proposed to facilitate the efficient completion of data analysis tasks. Visualization of spatial properties is often map-based~\cite{ferreira2013visual, lu2016exploring}, such as heat maps. 
Rendering and aggregation are effective solutions to visual clutter~\cite{zhou2013edge, feng2020topology}. 
For temporal properties, axis-based design is a common visual form~\cite{ferreira2013visual}. ThemeRiver~\cite{guo2011tripvista} and horizon map~\cite{suh2019persistent} can compare multiple properties over time. What's more, calendar~\cite{silva2019predicting} and radial layout~\cite{zhou2018visual, al2016semantictraj} can be a good option for periodicity, which can provide a distinct comparison of different periods. 
Additionally, a good comparison design is essential to grasp the difference. Juxtaposition is the most commonly used method for comparison~\cite{8017655, liu2016smartadp, weng2020towards}. Superposition overlaps multiple objects to show the difference~\cite{ferreira2013visual}. Explicit encoding presents differences visually. For instance, the grid heat map encoded by size and color in~\cite{8691491} showed the pattern similarity between different time periods.
At last, interaction operation helps with combining human knowledge. Basic interactions like clicking and boxing can promote exploration from overview to detail, such as the flow matrix view in~\cite{weng2020towards}, which can be clicked for more information. What's more, interactions such as sketch~\cite{8691491} can make exploration more interesting, enabling people to interact with systems.

Although these studies have been proven effective, they cannot directly used for our visual analytics tasks. 
For the taxi data used in this paper, relatively little research has been done on pattern analysis of passengers' behavior for pick-up selection. It is still a challenge to explore enormous pick-up points. Problems such as visual redundancy and poor scalability are easy to appear. 
Therefore, we combined the multiple visualization techniques and designed a novel visual analytics system, which enables users to interactively explore pick-up selection behavior at multiple scales. In particular, a juxtaposition comparison view was designed to interact with plenty of pick-up points and gathering locations, depicting different patterns in different regions.

\section{Data and Analytical Tasks}
\label{sec:3}

This section first describes the data processing procedures and the derived output. Next, we derive a list of analytical tasks by working with domain experts.


\subsection{Data Description}
In this research, {\name} is constructed from three types of data. 
i) We utilize the road network of Shenzhen and geographical data from the open street map\footnote{\href{https://www.openstreetmap.org/}{https://www.openstreetmap.org/}}. 
ii) We retrieve Points of Interest (POIs) data from the Baidu Map Service\footnote{\href{https://api.map.baidu.com/lbsapi/getpoint/index.html}{https://api.map.baidu.com/lbsapi/getpoint/index.html}}. The dataset includes 190362 POI locations where each record contains the longitude, latitude, name, address, and functionality of a structure in the urban environment. 
iii) We use the taxi GPS data from September 1, 2019 to October 5, 2019 in Shenzhen city. The raw data records the location of each taxi with a total size of over 200 GB. Each record consists of a timestamp, taxi id, longitude, latitude, speed, travel direction, and occupancy status. 

\subsection{Data Processing}
\label{sec:dataProcessing}
\begin{itemize}
\item \textbf{POI classification:}
The raw POI data is categorized into 20 industry categories and 158 detail categories. 
To better focus on the POI information related to our scenario, we reclassified POI data into six categories: company, education (i.e., schools and universities), entertainment (i.e., restaurants, shopping malls, and tourism), living, public service (i.e., government agency) and traffic(i.e., station and parking lot).

\item \textbf{Trip extraction:}
Taxi GPS data records a sequence of key points that a taxi passed by, which is unclear for the pick-up point (Origin) and drop-off point (Destination) of a single trip. So we need to extract each OD trip from raw GPS data. 
It consists of three steps: i) Data cleaning. Some records that are out of the study area or occupancy status changes abnormally, are eliminated at first. ii) Pick-up and drop-off points identification. It can be extracted according to the change in occupancy status. It means that someone gets off when the taxi's status changes from 1 to 0, the opposite means someone getting on. iii) \gsx{Attribute addition}. 
\gsx{To facilitate the analysis of pick-up point selection with surrounding POIs,} we attach the category attribute of the nearest POI to each OD point. On average, about two million trips are extracted each day. The metadata of each trip record is shown in Table~\ref{table:OD}.

\begin{table}
\caption{\gsx{The Metadata of Each OD Trip Record}.}
\begin{center}
\begin{tabular}{|l|c|}
\hline
Field & Description \\
\hline\hline
ID & The id of the taxi \\ 
\hline
stime & Trip start time \\
\hline
slocation & Location of pick-up point \\
\hline
etime & Trip end time \\
\hline
elocation & Location of drop-off point \\
\hline
cO & Category of the nearest POI of the O point \\
\hline
cD & Category of the nearest POI of the D point \\
\hline
\end{tabular}
\end{center}
\vspace{-7mm}
\label{table:OD}
\end{table}

\item \textbf{Geographical partition:}
In order to facilitate pattern exploration, we divide the study area into equal-sized grids. Firstly we consider the shape of the grid. 
\gsx{Hexagon is chosen because it is recommended as a better alternative to be used as a statistical unit~\cite{rempel2003patch}. The main advantages are i) that hexagons are the only geometric shape for regular tessellations that shares a real border with every neighbor and not only a single point with some neighbors~\cite{adamczyk2017zonalmetrics} and ii) hexagons have minimal visual ambiguity, and have a positive effect on the memory performance of object location~\cite{edler2019hexagonal}.
Then we tested the grid at different resolutions (200 m, 400 m and 1 km), finding that 400-meter hexagons provide representative samples which can guarantee the details of the pick-up point distribution and is in line with the passenger's willingness to walk the distance.
After partition, we assign taxi movements to each hexagon based on their original location.}

\end{itemize}

\vspace{2mm}
\subsection{Task Analysis}
\label{sec:analyticsTasks}
To develop a feasible and practical approach for analyzing and improving pick-up points with visual analytics, we
work closely
with our collaborative experts (E1, E2, E3) to derive the analytical tasks.
E1 is a Ph.D. candidate specializing in urban visualization and is also one of the co-authors.
E2 is an analyst in a taxi company with over 10 years of working experience and has been involved in several taxi management projects.
E3 is a researcher who has long been engaged in urban computing and visual analytics.
From the feedback of these experts, we summarize a set of system requirements as shown below:

\begin{itemize}
\item [\textbf{T1}]
\textbf{Obtain an overview of global traffic distribution in a city.} The users need to grasp the spatio-temporal distribution of pick-up points over the city. A summary of daily traffic volumes provides information for time checking (T.1.1). Then the distribution of pick-ups in different regions helps to understand busy regions for further exploration (T.1.2).

\item [\textbf{T2}]
\textbf{Provide the spatio-temporal patterns for regional traffic.} After understanding the global distribution of pick-up points, users need regional analysis. Multiscale temporal pattern at the hourly, daily, and weekly periods is noteworthy (T.2.1). Besides, it is necessary to explore the traffic within a region (T.2.2), such as the traffic volumes of pick-ups and drop-offs, and the direction of traffic. E2 points out that drop-offs are also important information, meaning that there are potential passengers subsequently or more drivers are here. These help the driver make decisions about cruising.

\item [\textbf{T3}]
\textbf{Compare the behavior of pick-up point selection with different regions.} Visual comparison of situations across different regions should be supported. It is necessary to compare different locations diversely to find distribution similarities and differences, facilitating the discovery of preference patterns for pick-up point selection.

\item [\textbf{T4}]
\textbf{Show detailed information for point analysis.} In a focused region, users expect to explore the preference of passengers in choosing different pick-up points. A set of criteria is necessary to explore and compare the performance of different pick-up points.

\end{itemize}


\begin{figure}[t]
\begin{center}
   \includegraphics[width=0.9\linewidth]{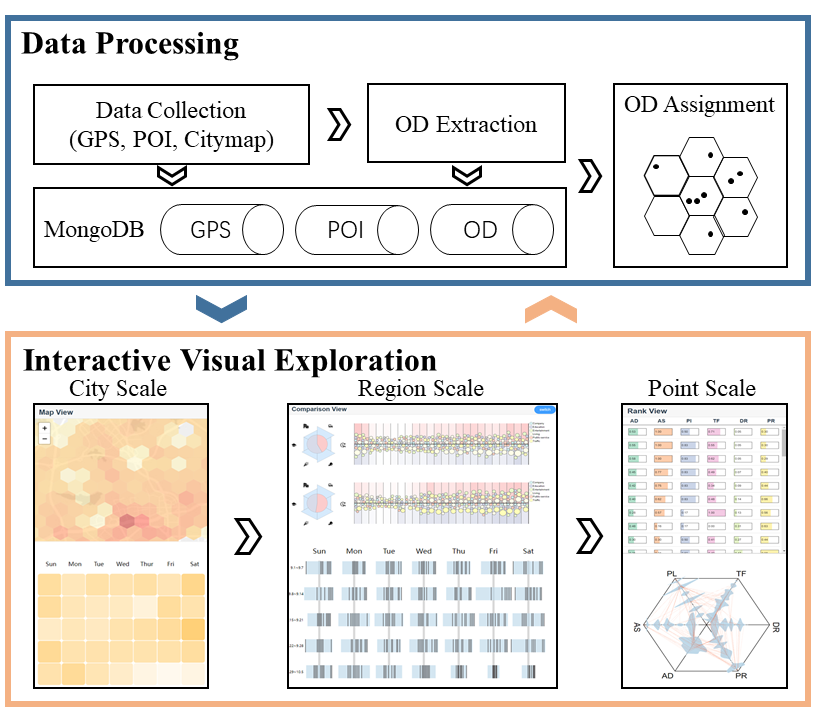}
\end{center}
\vspace{-5mm}
   \caption{A visualization system pipeline for multi-scale analysis of behavior for pick-up point selection. Our system consists of two phases: data processing and interactive visual exploration. In the data processing phase, we perform well-established methods to process data and index them spatially in the database. In the interactive visual exploration phase, four coordinated views are provided to support three-scale exploration.  }
\label{fig:pipeline}
  \vspace{-4mm}
\end{figure}

\section{System Overview}
\label{sec:4}

\name~is a web-based visual analytics application constituted of two phases, namely, data processing and interactive visual exploration, as illustrated in \autoref{fig:pipeline}. In the data processing phase, {\name} processes datasets offline and stores them in the MongoDB database. We extract OD trips from raw GPS data and index them spatially in the database. Then we divide the study area into grids and assign OD trips to each grid based on location.

The interactive visual exploration phase consists of three stages of multi-scale exploration. We organize the interface by the city-, region-, and point-scale analyses. Starting from the city scale, a heatmap in map view (\autoref{fig:overview}$\rm B_1$) and a calendar chart (\autoref{fig:overview}$\rm A_3$) are designed to provide a spatio-temporal overview of city traffic distribution (\textbf{T1}). Users can have a basic understanding about data. Then narrowing down the exploration to region scale, we design a multiscale temporal chart for regional exploration (\textbf{T2}) and a comparison view (\autoref{fig:overview}C) to compare region patterns in different situations (\textbf{T3}). At last, users can analyze the preference for pick-up point selection at point scale. 
A rank view (\autoref{fig:overview}$\rm D$) is provided to compare different points and summarize their performance (\textbf{T4}).
The visualization modules are implemented in D3.js\footnote{\href{https://d3js.org/}{https://d3js.org/}} and Leaflet.js\footnote{\href{https://leafletjs.com/}{https://leafletjs.com/}} for different rendering requirements, and they are integrated using Vue.js\footnote{\href{https://vuejs.org/}{https://vuejs.org/}}framework.

\section{Visualization}
\label{sec:visualization}
In this section, we provide a detailed description of the visualization designs in our system.







\subsection{\temporalViewCapital}
The {\temporalView} (\autoref{fig:overview}A) is designed to configure the system and provide multi-scale temporal information (\textbf{T1, T2}). 
At the top (\autoref{fig:overview}$\rm A_1$), a date selector is provided for users to select traffic on a particular day they are interested in. Then a view selector (\autoref{fig:overview}$\rm A_2$) is provided to configure the visibility of different views flexibly through a set of checkboxes.
To obtain an overview of global traffic distribution over time for date selection (\textbf{T.1.1}), a calendar heatmap (\autoref{fig:overview}$\rm A_3$) is designed to summarize the daily traffic volume of the whole city. Each rectangle represents a day and the color is encoded as the volume of the daily traffic. The darker rectangle indicates the larger traffic volume in a day.

\begin{figure}[t]
\begin{center}
   \includegraphics[width=0.9\linewidth]{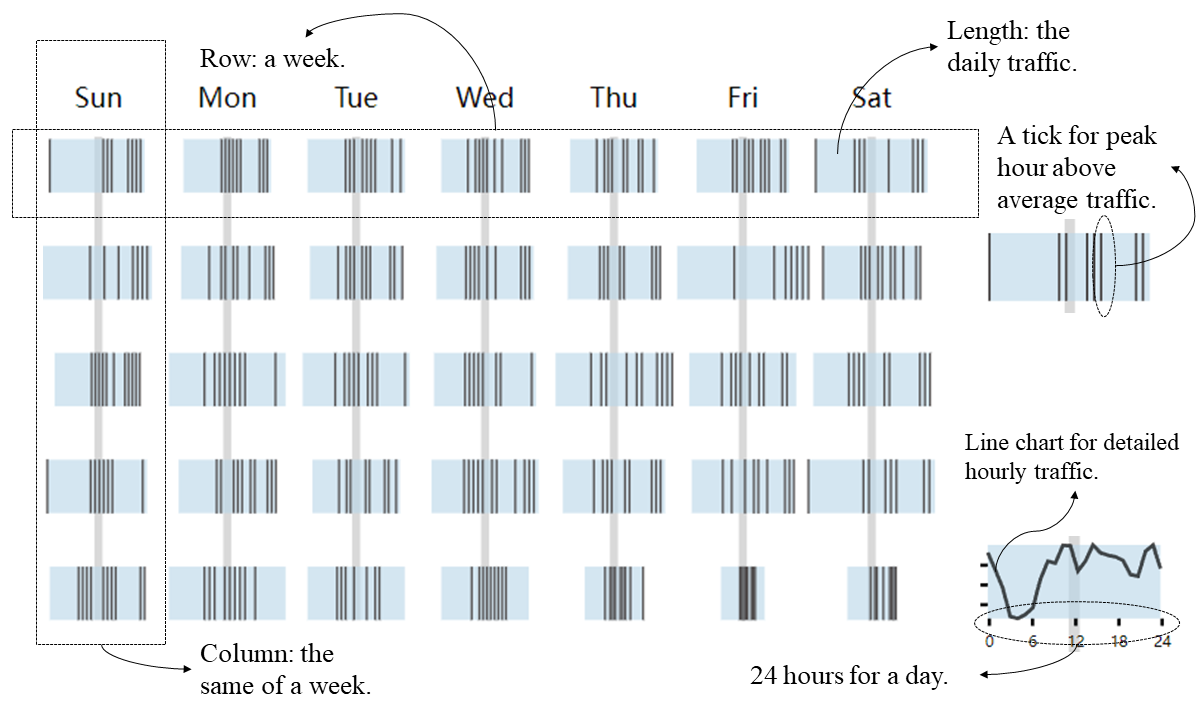}
\end{center}
\vspace{-6mm}
   \caption{The design of the multiscale temporal chart, shows temporal patterns at different granularities. \gsx{Each row represents a week, and each column represents the same day of the week.} It supports traffic volume comparison weekly, daily, and hourly. }
\label{fig:barchart}
  \vspace{-3mm}
\end{figure}

At the bottom of the {\temporalView} (\autoref{fig:overview}$\rm A_4$),
we design a multiscale temporal pattern chart
to better reveal the temporal distribution of pick-up points in the focused region (\textbf{T.2.1}).
It shows changes of traffic at different time granularities. \autoref{fig:barchart} shows the design details. The layout of blue rectangles is similar to the calendar heatmap, with each row representing a week, and each column representing the same day of the week. 
For each rectangle, we use length to show the daily traffic volume. 
\gsx{Furthermore, the hourly pattern is also worthy of attention. Users are concerned about changes in hourly traffic. First, we only highlight the peak hours with higher traffic than the average per day. We split each rectangle's length into 24 equal pieces to represent the 24 hours in a day. We encode these hours with ticks and placed them in each rectangle from left (00:00) to right (24:00). For better comparing and analyzing, the rectangles in each column are centered, with dashed gray lines indicating 12:00.}
Then E2 suggested that he would also like to know the details of how traffic flows change within one day, which ticks cannot show. So we provide a choice to check traffic changes hourly in a day with a line chart, which can be viewed by clicking on the rectangle. The graph is initialized to the traffic volume for the entire city and then changes with the selected region.

\subsection{\mapViewCapital}

The {\mapView} (\autoref{fig:overview}B) provides a spatial overview of pick-ups (\textbf{T.1.2}) and supporting regional exploration (\textbf{T.2.2}).

\textbf{Heatmep view.} Traffic in a large-scale city typically comprises massive trips. The heatmap (\autoref{fig:overview}$\rm B_1$) is developed to assist with mastering the global distribution of OD trips quickly and identifying regions of interest (\textbf{T.1.2}). It encodes the density of pick-ups at each grid. The darker, the higher number of pick-up points. From the heatmap, users can easily find crowds from a coarse-grained perspective. 

\begin{figure}[t]
\begin{center}
   \includegraphics[width=0.9\linewidth]{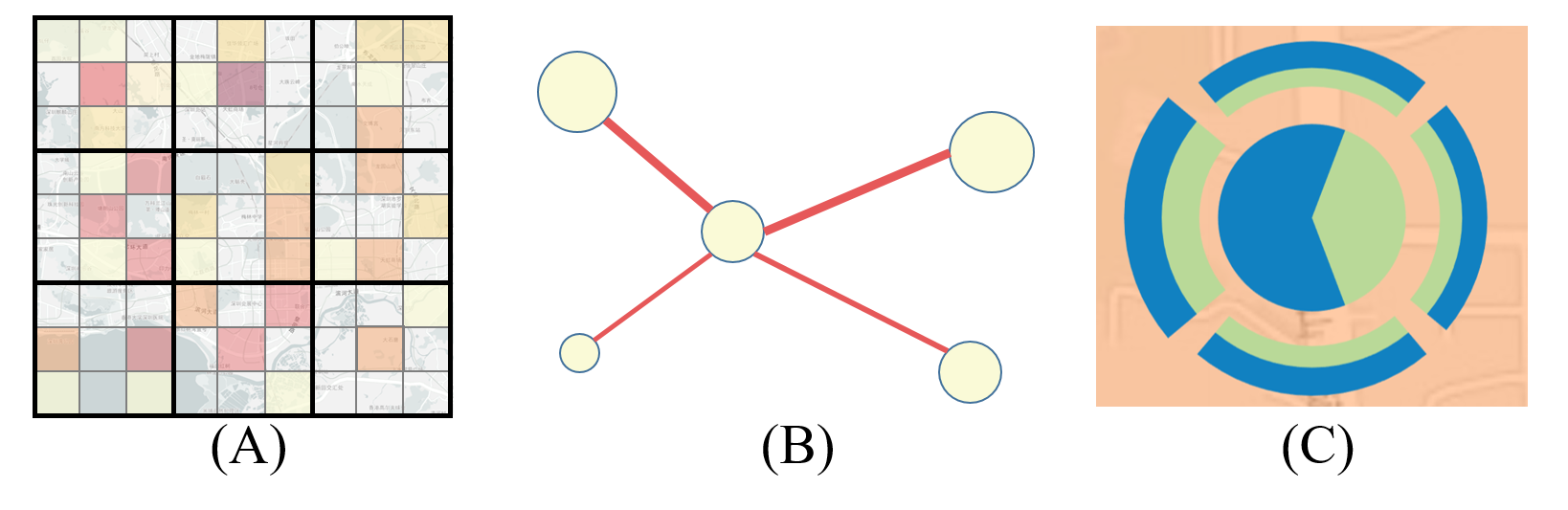}
\end{center}
\vspace{-5mm}
   \caption{Designs of the glyph view. (A) and (B) are the design alternatives of the glyph view, which may introduce visual clutter problems. (C) is the final adopted design.}
\label{fig:alternativeGlyph}
  \vspace{-5mm}
\end{figure}

\textbf{Glyph view.} To reveal traffic pattern in a focused region (\textbf{T.2.2}), we attach a glyph view (\autoref{fig:overview}$\rm B_2$) to the heatmap. 
As shown in \autoref{fig:alternativeGlyph} C, the pie chart inside represents the comparison between the pick-ups (blue) and drop-offs (green) of each grid. Two outer rings visualize the drop-offs (green) and pick-ups (blue) in the grid by the geographical directions, with thickness presenting the flow size. Users can select regions with the lasso tool (\autoref{fig:overview}$\rm B_4$).

\textbf{Design alternatives.}
We considered several alternative solutions during our glyph design process. We first tried the OD matrix~\cite{wood2010visualisation} to illustrate the movement of OD trips. As shown in \autoref{fig:alternativeGlyph}A, we tried to mesh the city map and map the original grids of the entire city into every single grid. The color was used to encode the volume of each grid area to other grids. But it has several drawbacks. First, it encodes the relative position of the destination, so it is difficult to fix the position of the destination intuitively. Second, the dense grid and colors make it hard to spot patterns. Moreover, we tried the flow map~\cite{von2015mobilitygraphs}  as shown in \autoref{fig:alternativeGlyph}B. However, it introduced a visual clutter problem when there were too many grids. So in the end we chose the glyph design (\autoref{fig:alternativeGlyph}C), which was also recognized by the experts. It can effectively display the pick-up and drop-off information and flow direction in different grids.

\textbf{Point view.} To assist visual linkage and smoother operations, we present a point view (\autoref{fig:overview}$\rm B_3$) to provide a spatial context for point analysis. Different types of POIs are marked on the map with various icons and a donut graph is placed around the POI to summarize the nearby pick-ups (blue) and drop-offs (green) for taxis. If users click the icon, the detailed information for the POI will be displayed (\autoref{fig:overview}$\rm B_5$). 

\subsection{\comparsionViewCapital}
To compare the behavior of pick-up point selection in different regions (\textbf{T3}), the comparison view (\autoref{fig:overview}C) coordinates two graphs for two regions at the same time.

\begin{figure}[t]
\begin{center}
   \includegraphics[width=0.9\linewidth]{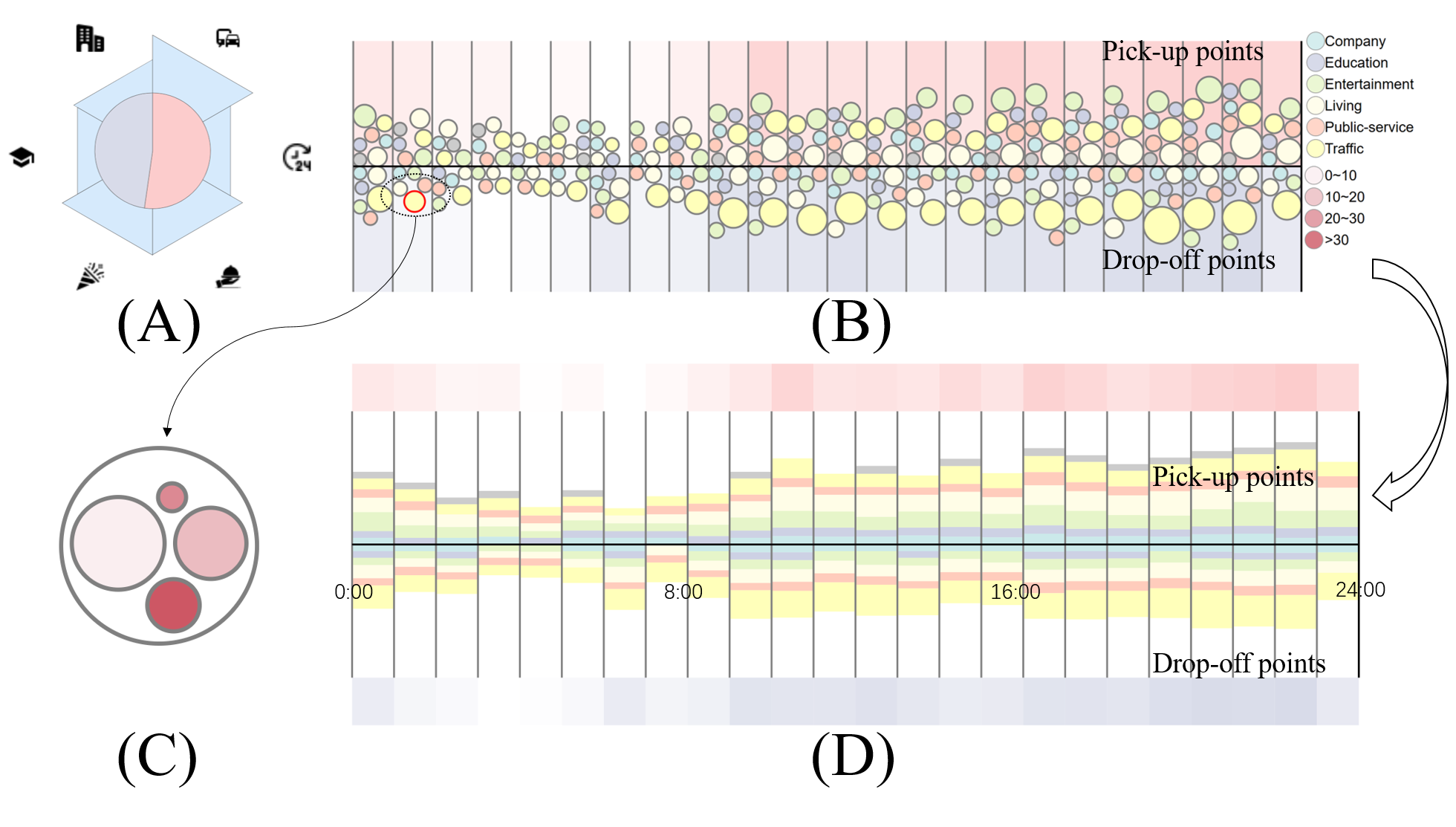}
\end{center}
\vspace{-5mm}
   \caption{Design details for the comparison view. A glyph shows the comparison of pick-ups and drop-offs and the overall distribution of POIs (A). Two kinds of options are provided to switch for details (B, D). A circle pack displays the information of trip duration (C). }
\label{fig:comparison}
  \vspace{-3mm}
\end{figure}

\autoref{fig:comparison} shows the design details. For each region, a glyph on the left (\autoref{fig:comparison}A) summarizes the information for the region. The middle pie chart displays the proportion of all-day pick-ups (pink) and drop-offs (purple) in the region. The outer arc bars represent different categories of POI with height encoding the number.
Afterward, users expect to explore the preference of pick-up selection, such as the preferred period and locations.  
Inspired by the 1-D beeswarm graph\footnote{\href{https://observablehq.com/@fil/experimental-plot-beeswarm}{https://observablehq.com/@fil/experimental-plot-beeswarm}}, we design a novel 2-D visual metaphor to represent the pattern in a region (\autoref{fig:comparison}B). 
As shown in \autoref{fig:comparison}B, the x-axis represents time, and the y-axis represents the pick-up or drop-off points. The background represents the pick-up (top) and drop-off (bottom) hourly.
Then we count the number of pick-ups and drop-offs by an hour and aggregate them according to their POI attributes, trying to present the time-varying flow and spatial properties of the pick-up point.
These points for each hour are encoded into circles with different colors indicating different POI attributes and placed in the corresponding grids. The area of the circle indicates the quantity.
\gsx{Furthermore, more detailed information about the corresponding trips is needed. Experts point out that drivers are concerned about the travel duration of an order. Therefore, a circle pack (\autoref{fig:comparison}C) is provided to summarize the travel duration by clicking on each circle.} We divide trips into four categories according to the time spent on the trip: 0-10 minutes, 10-20 minutes, 20-30 minutes, and more than 30 minutes. Each circle represents a category, with the area of the circle indicating the number of trips. The darker, the longer time spent on the trip. Then force simulation is used to keep all the circles tightly connected without overlapping for readability. \gsx{In this way, the beeswarm graph
can provide an overview of the temporal distribution and also enable detailed local inspection.
In addition, the number of circles is sometimes large, affecting the efficiency of the comparison analysis. With this in mind, we provide checkboxes on the right to support point filtering by POI attribute, facilitating focus on points of interest.} 

The new beeswarm graph we design can effectively display the spatio-temporal information of the region and help to interact with a large number of trips, facilitating pattern discovery. However, according to the feedback from E1, when the selected region is too large, the circles overflow occurs. So we provide the option of a stacked bar chart (\autoref{fig:comparison}D) for users to switch. Its layout is similar to the beeswarm graph, which can more clearly show the changes over time. The original background is pushed to the top and bottom.

\begin{figure}[t]
\begin{center}
   \includegraphics[width=0.9\linewidth]{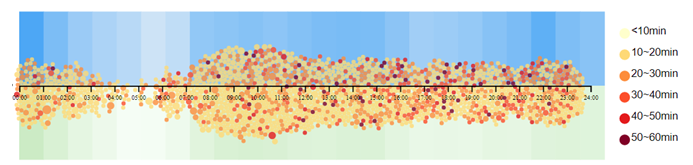}
\end{center}
\vspace{-4mm}
   \caption{Alternative design for the comparison view, which is difficult for users to interact with circles.
   }
\label{fig:alternativeComparison}
  \vspace{-3mm}
\end{figure}

\textbf{Design alternatives.}
For the beeswarm graph, we have considered aggregating data by the taxi ID to present hourly information, hoping to alleviate large-scale OD data. As shown in \autoref{fig:alternativeComparison}, we use each circle to represent a taxi, with the area of the circle representing the number of orders and the color showing the average travel duration.
However, the visual presentation is confusing due to the still large number of taxis. It is difficult for users to effectively interact with circles.
So we assign the POI attribute to each pick-up point according to the nearest POI. Then we aggregate the pick-up points according to the POI attributes, and distribute the pick-up points and the drop-off points separately on the positive and negative axes of y, which effectively solves this problem. At the same time, it provides an opportunity to explore the spatial characteristics of the pick-up points.

\subsection{\rankViewCapital} 
After obtaining the patterns of the region, point-scale exploration (\textbf{T4}) is necessary to compare the performance of each pick-up point and explore the preference for passengers in different pick-up points. We design the {\rankView} (\autoref{fig:overview}D) to compare and analyze pick-up points, which can help with driver cruise guidance.
In reality, POIs are usually recommended as pick-up points because they are easy to find. Considering the large number of historical pick-ups, as well as the habit of passengers taking a taxi, we use POI to replace historical pick-ups for analysis. Users can also add points on the map as a supplement. They are collectively called candidate points.

Firstly, a set of criteria is required to evaluate the performance of these points. According to literature research~\cite{chen2020finding,7000580,zhu2019pick} and expert opinions, we propose a set of criteria.
Our criteria are as follows.
For each candidate point i: i) AD: Accessibility is assessed by the average distance from nearby historical points within coverage to that point. $D_{ij}$ in Formula (\ref{formula:AD}) means the distance from the historical point j to i. $D$ is the radius of the coverage area. $n$ is the total number of historical points. Finally, candidate points with better accessibility have higher scores.
ii) AS: Average traffic speed in the vicinity represents the traffic smoothness near the candidate point. A high score indicates a safe and comfortable ride experience.
iii) PL: POI level is the proportion of nearby POI categories in all, indicating the convenience of taking a taxi. 
iv) TF: Transfer convenience, impacting passenger source, means the proportion of transportation facilities in all POIs.
v) PR and vi) DR evaluate the probability of passenger arrival and driver discovery respectively. In Formula (\ref{formula:PRDR}), $NP_i$ and $ND_i$ mean the number of historical pick-up points and empty taxis. Then we calculate the quantity per unit length and unit time.
Finally, we normalize all scores to [0, 1].

\begin{equation}
    AD_i=\sum\limits_{0<j<n}(1-D_{ij}/D)/n
    \label{formula:AD}
\end{equation}

\begin{equation}
PR_i=NP_i/L/T, DR_i=ND_i/L/T 
\label{formula:PRDR}
\end{equation}

Afterward, a rank list (\autoref{fig:overview}$\rm D_1$) with six evaluation criteria is designed for comparing candidate points. The list presents each candidate point in a row and arranges its criteria in six columns. We encode the scores by the length of bars and provide a click action on the header of each column to rank the points by different criteria. 
Finally a radar graph (\autoref{fig:overview}$\rm D_2$) is placed under the list to provide a summary depiction of all candidate points. Each axis from the center to the outer edge represents a score from 0 to 1. The scores for each candidate point are connected by light orange lines. And the violin plots colored in blue are attached to each axis, showing the score distribution of all candidate points for that criterion, which emphasizes the pick-up preference. Users can click points of interest on the rank list. Then the scores of the corresponding points will be highlighted with thicker orange lines in the radar graph (\autoref{fig:overview}$\rm D_2$).

\subsection{User Interactions}
Rich interactions are provided for users to explore pick-up point data, which are summarized as follows.

\begin{itemize}
\item \textbf{Multi-scale navigation} helps users navigate effectively across different scales. In the {\temporalView}, users can set date (\autoref{fig:overview}$\rm A_1$) and configure the visibility of views flexibly through a set of checkboxes (\autoref{fig:overview}$\rm A_2$).
In the {\mapView}, users can select regions with the lasso tool (\autoref{fig:overview}$\rm B_4$) or set a point by clicking on the map (\autoref{fig:overview}$\rm B_6$). 


\item \textbf{Highlighting} enables users to focus on the information of interest, which is supported in the {\mapView} and {\rankView}. For example, the selected region and point will be highlighted in the {\mapView} (\autoref{fig:overview}$\rm B_4$, $\rm B_6$) and the {\rankView} (\autoref{fig:overview}$\rm D_2$).

\item \textbf{Linking} connects four views in the system. For example, after setting points on the map, new points will be added at the end of the rank list. When users click on a row in the rank list, it can be visually linked to the location on the map and the score on the radar chart.
\end{itemize}

\section{Evaluation}
\label{sec:evaluation}

In this section, we demonstrate the effectiveness and usability of {\name} to accomplish the visualization tasks in Section~\ref{sec:analyticsTasks} and discover insights through three case studies and expert interviews with the aforementioned collaborating experts (E1, E2 and E3, who have been introduced in Section 3.3). 
The dataset used in these three case studies is a 5-week taxi GPS dataset in Shenzhen, from September 1st to October 5th, 2019.

\subsection{Case study}


\subsubsection{\gsx{Taxi pick-up behavior comparison across different regions}}

\label{sub:case1}

\begin{figure}[t]
\begin{center}
   \includegraphics[width=0.9\linewidth]{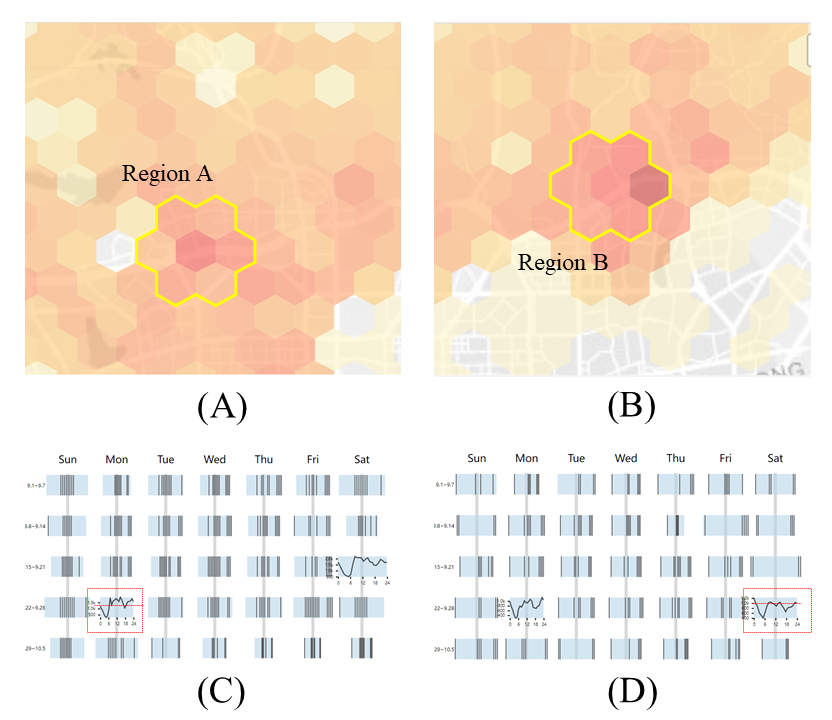}
\end{center}
\vspace{-7mm}
   \caption{(A) and (B) show the selected regions for pattern comparison. (C) and (D) are the corresponding temporal patterns over a month in region A and B. The traffic in region A is busy during the day while the traffic in region B is busy at night. }
\label{fig:region1}
  \vspace{-5mm}
\end{figure}

As an expert in urban visualization, E1 was interested in using our system to explore the pick-up point selection patterns in different regions of the city.
After loading data to {\name}, the temporal overview of pick-ups across the city was displayed in the calendar heatmap (\autoref{fig:overview}$\rm A_3$) first. E1 observed darker rectangles on weekends than weekdays, indicating higher traffic volume on weekends. 

\begin{figure*}[!t]
 \centering
 \includegraphics[width=1.88\columnwidth]{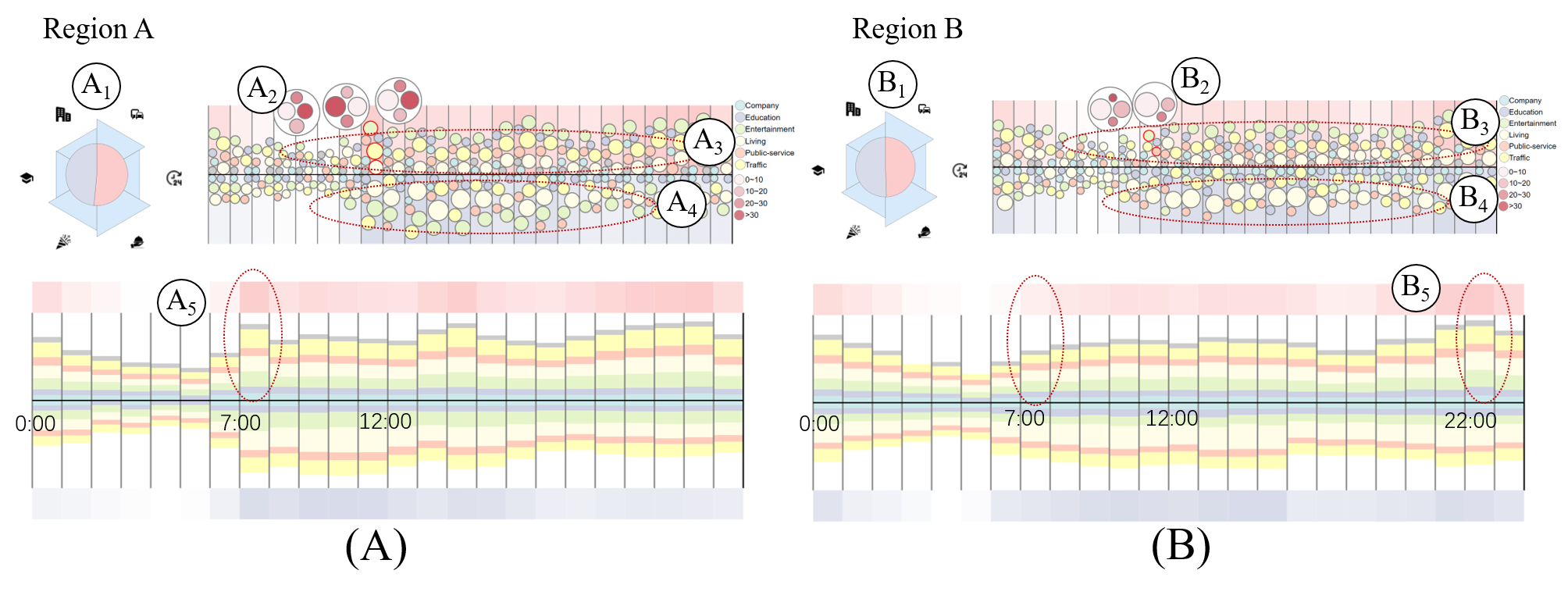}
  \vspace{-5mm}
  \caption{Pattern comparison between two selected regions. (A) shows two presentations of region A on a weekday. Region A has more long-duration trips in the morning and more trips for entertainment. (B) provides two graphs of region B on a weekday. Region B has obvious characteristics of the evening peak.}
\label{fig:case1}
\end{figure*}

He chose a Monday (September 23, highlighted with a blue rectangle in \autoref{fig:overview}$\rm A_3$) and immediately found two hotspot regions (\autoref{fig:region1}A, B) on the map view. After checking, E1 found that they were both in the downtown area, with massive traffic. After selecting these regions with the lasso tool, he observed the changes in the multiscale temporal chart (\autoref{fig:region1}C, D). For region A, \autoref{fig:region1}C showed that the ticks were concentrated in the middle, indicating more peak hours in the daytime. While the ticks in \autoref{fig:region1}D were distributed on both sides, meaning more traffic in the evening in region B.
Afterwards he examined the comparison view (\autoref{fig:case1}) to compare two regions. The left glyphs for two regions (\autoref{fig:case1}$\rm A_1$, $\rm B_1$) were similar, showing that they had similar distributions of POIs with roughly equal pick-ups and drop-offs. In the right beeswarm graphs for two regions, he found that the various circles above the x-axis (\autoref{fig:case1}$\rm A_3$, $\rm B_3$) were similar in size. It was hard to for E1 to find passengers' preference in specific POI categories when they choose the pick-up point. However, for the circles below the x-axis (\autoref{fig:case1}$\rm A_4$, $\rm B_4$),  the white and green circles were significantly larger than others, indicating the number of drop-offs near ``Living" and ``Entertainment" was increased. ``Living" and ``Entertainment" POIs were the preferred drop-off points. 
E1 thought that the drop-off points could strongly reflect the purpose of the trip, which may be the pick-up point of the next trip. ``It can provide useful suggestions for driver cruise," said E1.

Next, he switched to the stacked bar chart (\autoref{fig:case1}$\rm A_5$, $\rm B_5$) to explore time variation of traffic. This time he focused on pick-ups above the x-axis. He observed the higher bar and darker rectangle at 7:00 AM in \autoref{fig:case1}$\rm A_5$, showing a more obvious peak hour at 7:00 AM than region B. 
He felt interested in the rush hour at 7:00 AM. To further explore trips at 7:00 AM, he switched back to the beeswarm graph and checked interested circles. Comparing \autoref{fig:case1}$\rm A_2$ and \autoref{fig:case1}$\rm B_2$, the dark red circles of circle packs in \autoref{fig:case1}$\rm A_2$ were larger than others, meaning more long-duration trips ($>$30min) at 7:00 AM in region A than B. In contrast, short-duration travel accounted for the main part in region B (\autoref{fig:case1}$\rm B_2$). 
Learned from E1, there were many people need to travel a long distance to go to work in region A. Although there were plenty of pick-up points in both areas, ``region A has a huge demand for picking up in the morning, and longer trips bring higher incomes for drivers," said E1.


The above research confirmed that {\name} can effectively help to explore pick-up selection patterns in different regions. E1 said, ``It is effective to support exploration at multiple scales, which helps a lot for taxi dispatch."

\subsubsection{
Taxi pick-up point selection in the living region}
\label{sub:case2}

As an analyst in a taxi company, E2 works on taxi dispatch. He was interested in the characteristics of passengers in selecting pick-up points, wondering if there are any regularities and preferences.

\begin{figure}[t]
\begin{center}
   \includegraphics[width=0.9\linewidth]{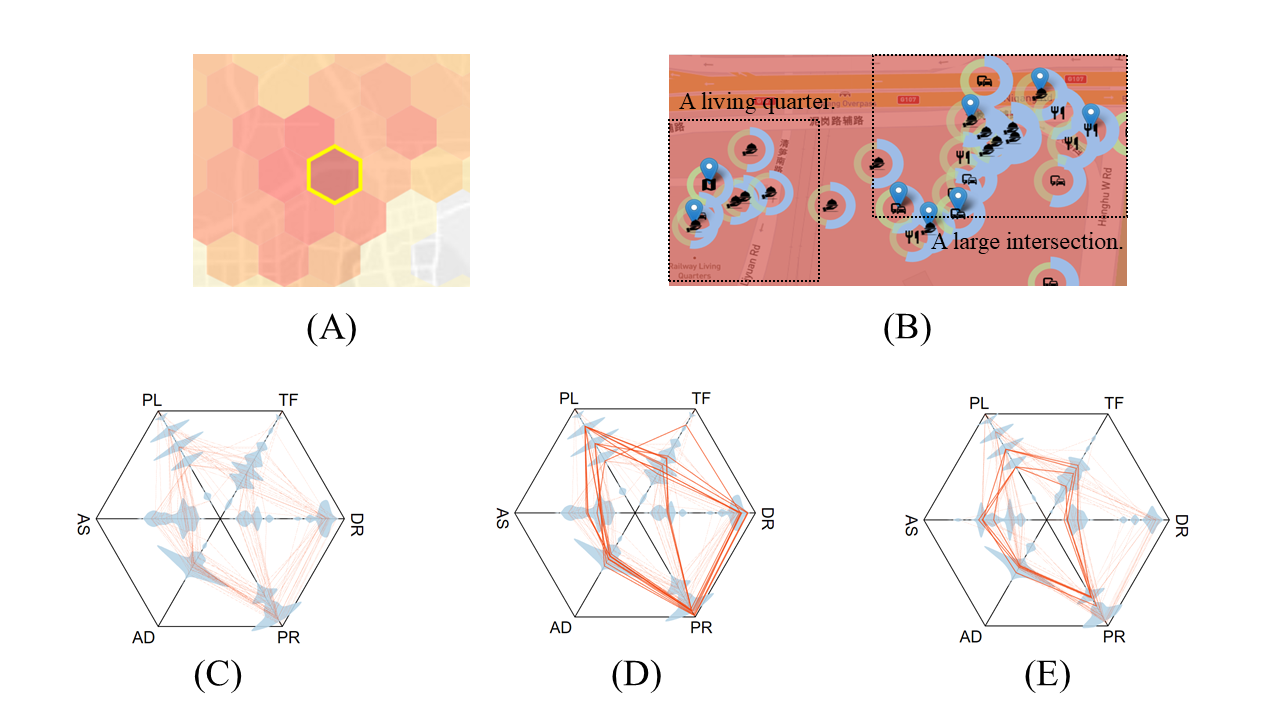}
\end{center}
\vspace{-7mm}
   \caption{Exploration for preference of pick-up point selection in a living quarter. (A) shows the selected region in Luohu District. (B) shows that there are more pick-ups at the larger intersection. (C) displays the overall performance of all candidate points. (D) and (E) display the performance of selected points. Passengers prefer to pick up at POI-rich places.}
\label{fig:case2}
  \vspace{-3mm}
\end{figure}

E2 checked the calendar heatmap and selected a normal weekday (September 23) for exploration. He quickly discovered the dark red regions with high traffic volume in Shenzhen. For the large amount of taxi demand, E2 was eager to understand the preference of pick-up point selection to provide detailed suggestions for taxi dispatch. He selected the darkest grid (\autoref{fig:case2}A), which contained a large living quarter. 
After clicking the ``point view" in \autoref{fig:overview}$\rm A_1$, the map view immediately zoomed to the point scale and the rank view ranked all POIs automatically. At E2's suggestion, we set D in Formula (\ref{formula:AD}) to be 500 meters to measure the pick-up situation near POI.
He checked the surrounding conditions of POI in the map view firstly. As shown in \autoref{fig:case2}B, the blue rings at the intersection were obviously thicker than those around the living quarter. It indicated that the pick-ups were concentrated at the further intersection, where there were abundant POIs. As he had thought there would be more pick-ups near the living quarter, he was surprised by the results. With doubt, he continued to inspect the rank view, hoping for an explanation.

As shown in \autoref{fig:case2}C, the violin plots (blue) of PL, TF, and DR were distributed on the outer edges of the radar chart. In summary, most of the pick-ups in the region had high scores and good performance. 
Afterward, he ranked all points by PR score (passenger arrival rate). He clicked the rank list and checked points with high PR scores, with corresponding scores highlighted in the radar chart with thicker orange lines (\autoref{fig:case2}D). These points had high PL, TF and DR scores, showing that most passengers preferred to wait for taxis in areas with rich POIs and convenient transportation. E2 said, ``Passengers think that the probability of taking a taxi is higher here. In fact, there are indeed more taxis passing by here." 
However, the lower AS (average speed) and AD (average distance) scores meant that it was easy to get congested and passengers had to walk for a long distance. 
Therefore, E2 ranked the points according to the AD scores from high to low and clicked on points with high AD scores for further exploration. The scores of these points were highlighted in \autoref{fig:case2}E. He found that some of these points had lower DR and PL scores, indicating fewer empty taxis and POIs nearby. The probability of taking a taxi at the nearby points was lower. Therefore, ``passengers have to walk a certain distance to the further intersection to take a taxi, which brings a bad ride experience," said E2. It resulted that the pick-up points were clustered at the further intersection rather than the living quarter. He suggested, ``to improve passengers' riding experience, it is better to recommend drivers to cruise around the living quarter, which also benefits traffic management."

E2 praised, ``{\name} is helpful for my work. It helps to identify abnormal areas quickly and provide information to make decisions."

\begin{figure}[t]
\begin{center}
   \includegraphics[width=1\linewidth]{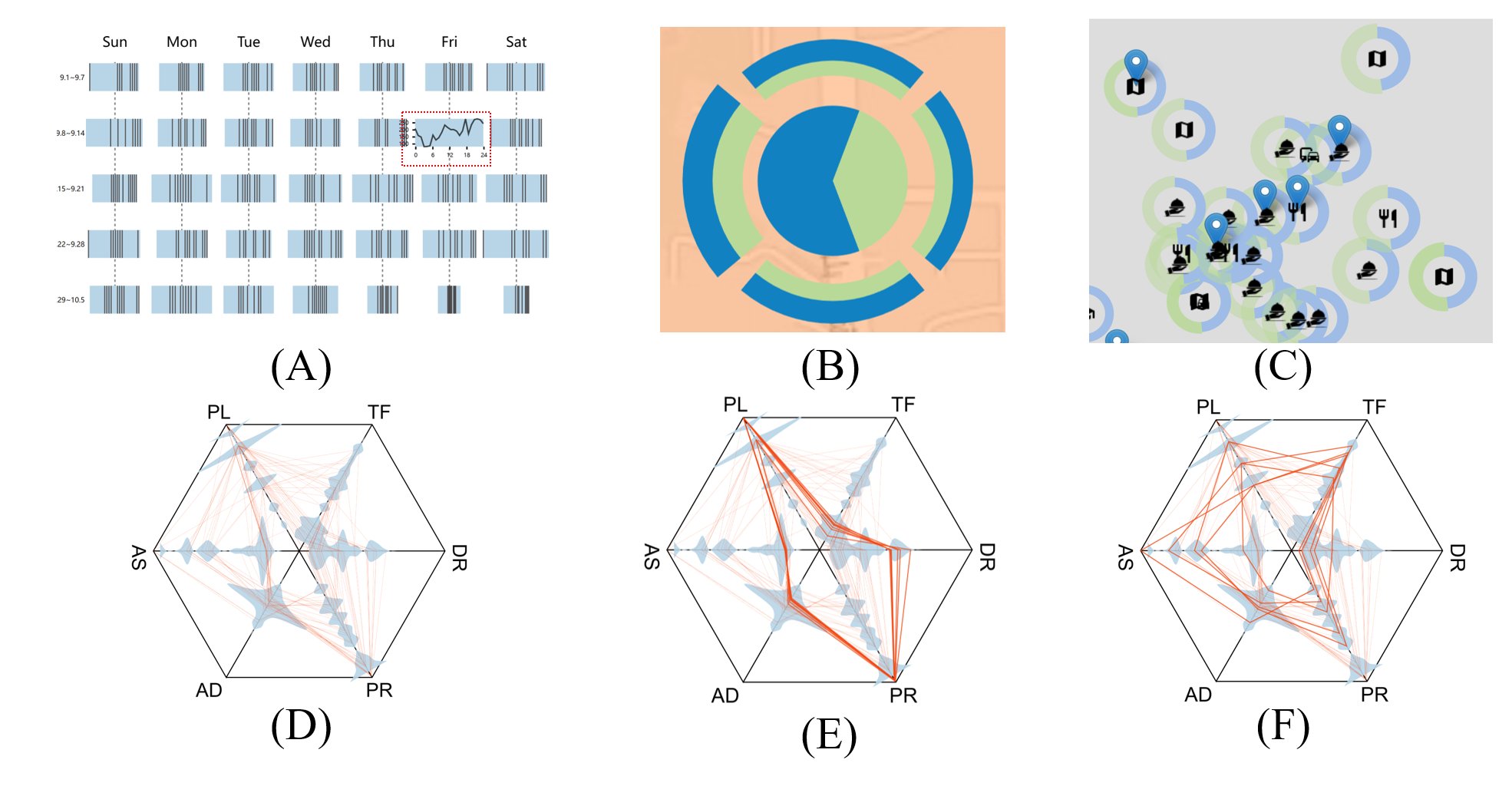}
\end{center}
\vspace{-8mm}
   \caption{Exploration for preference of pick-up point selection in a tourist region. (A) shows the change of traffic volume in the region over a month. Rush hours are concentrated in the afternoon and evening. (B) indicates that more traffic comes from and goes to the western regions. (C) shows that there are more drop-off points than pick-ups. (D) displays the score distribution of all points. (E) and (F) shows scores of selected points. There are fewer pick-ups in places with high TF score because there are lots of parking lots and passengers don't like to take taxis here.}
\label{fig:case2.2}
  \vspace{-4mm}
\end{figure}

\subsubsection{
Pick-up point selection around tourist spots}
E3 was interested in the problem of taking a taxi in scenic spots. Therefore, he selected a holiday (September 13, highlighted with a green rectangle in \autoref{fig:overview}$\rm A_3$) and found a tourist region in the west of the city (\autoref{fig:overview}$\rm B_4$), with many parks, gymnasiums, and amusement parks. He first inspected the regional pattern. As shown in \autoref{fig:case2.2}A, pick-up points started to gradually increase in the afternoon. In \autoref{fig:case2.2}B, the rings on the left are thicker than the others, indicating that there were more trips from and to the western regions, where there were many apartments and universities. ``There may be potential passengers there." he said.

Then he switched to the point view for further exploration. He found that the green rings are thicker than blue ones (\autoref{fig:case2.2}C), indicating more drop-offs than pick-ups around POIs in this region. Then he examined the radar chart (\autoref{fig:case2.2}D) and found the violin plot of DR (driver discovery rate) was lower than PR (passenger arrival rate), meaning the number of taxis cannot meet the demand of passengers. To further explore the demand of passengers, he checked points with high PR scores, hoping to find passengers' preferences for pick-up selection. As highlighted in thick orange lines (\autoref{fig:case2.2}E), he observed that the PL (POI level) scores of these points were high while TF (transportation facility level) scores were low. It indicated that passengers preferred pick-up points where there were many POIs but few transportation facilities. 
To find out the reason, he then ranked all points according to the TF score from high to low and clicked on them one by one. As shown in \autoref{fig:case2.2}F, some points with high TF had low DR scores, meaning there were few empty taxis near these points. Afterward, he examined the corresponding points on the map view and found that many of them were parking spots (\autoref{fig:overview}$\rm B_5$). 
E3 guessed that self-driving tours may account for a certain proportion. This is the reason why drivers are unwilling to cruise here.
But in fact, there is still unmet demand for taxis. For unmet passengers' needs, it was inappropriate to recommend remote locations to passengers. ``It is better to create a good cruising environment for taxi drivers.
The nearest POIs in the tourist spot are easier to find and have the shortest walking distance for passengers."
E3 suggested. 

In summary, E3 found it interesting to explore preferences of pick-up point selection with our system, {\name}. He could easily identify factors that passengers are concerned about and find current deficiencies. He said, ``{\name} can be of great help in providing suggestions for better taxi service."

\subsection{Expert Interview}
\label{sub:interview}

We interviewed three aforementioned experts (E1, E2, and E3) individually and collected their feedback. Each interview lasted about an hour. First, a fifteen-minute tutorial was provided to participants, which outlined the functions, visual designs, and interactions of our system. Then, participants were allowed to freely explore provided data with our system for about forty minutes. After that, we collected participants' feedback and suggestions. The feedback from the experts was valuable based on their expertise, which is summarized as follows.

\textbf{Visual designs.}
The experts confirmed that our system is well-designed and could be easily understood by users with different backgrounds. In particular, E1 praised the {\comparsionView}. `` It is of great help to discover the spatial property characteristics of pick-up points," said E1. E3 favors the multiscale temporal chart. He said, `` It provides abundant temporal patterns at different granularity, which is helpful."

\textbf{Usability \& Effectiveness}.
All three experts agreed that our proposed method is helpful and effective to analyze the pick-up points selection of passengers.
E1 mentioned, ``This approach supports multi-scale exploration including the city, region, and point scale. The hierarchical analysis inspires my research, and the comparison view does really help in comparing patterns of selected regions.
E2 praised that {\name} is helpful to make use of large amounts of GPS data collected from taxis. ``{\name} is an effective tool that can provide an intuitive visualization, this improves the decision interpretation. Compared to the current tools, it allows me to easily discover more detailed information to guide drivers on the cruise," he commented. 
E3 said that ``{\name} provides visualization designs which are also easy to follow. It is friendly to the analysts lacking professional skills in statistics and machine learning methods." 

\textbf{Suggestions}.
During the interview, three experts also provided fruitful suggestions on improving {\name}. Both E1 and E2 mentioned that the region segmentation function in the map view still needs more improvement. In the current version, {\name} uses a hexagonal grid to segment the urban area based on the latitude and longitude, E1 and E2 suggested that it would be better to consider the spatial contextual information during the region segmentation. For example, they hope that a building or a unit (park, campus) can be assigned to a grid, which makes sense. Instead of a unit being divided into several parts, each part is in a separate grid. E3 said that it would be more effective if {\name} integrated more recommendation approaches like mainstream deep learning models. In addition, he suggested that {\name} could compare these different recommendation methods and give the corresponding optimal method under different scenarios.

\section{Discussion}
\label{sec:discussion}

In this section, we discuss lessons learned, the implications, and identify the limitations of our system and propose directions for future studies.


\textbf{Lesson learned.}
We have learned three lessons. 
First, visual analytics is helpful to analyze pick-up point selection behavior in such big taxi data. Generally, research mainly focuses on regional analysis or pick-up point recommendation. 
In this paper, we apply visual analytics to analyze passengers' behavior for pick-up point selection at multiple scales. By working with domain experts, we find visual analytics is especially helpful for comparing and reasoning different behavior of pick-up point selection. We can easily find some parts interesting or abnormal, and conduct further exploration. 
Second, to analyze such big spatial-temporal taxi data, a three-scale exploration strategy is effective. We start with the city scale, and then compare different regions, and further explore different pick-up points, which is smooth and helpful for exploration. 
\gsx{Third, extensive communication with end users is important in determining the system’s requirements and implementing the visualization forms. They provide useful insights that make our systems more responsive to the needs of real-world applications. Experts subscribe to the hierarchical analysis approach and help us refine our tasks. For instance, E2 states how the analysis of drop-off point is essential for assisting the driver cruise when exploring the pick-up point selection behavior. In addition, users are unfamiliar with visual analytics. We can only implement concise visualization designs and effective tools by communicating frequently.}

\textbf{Implications.} 
\gsx{This study presents a visual analytics system, assisting analysts in taxi companies to explore passengers' pick-up point selection behavior. In terms of taxi companies, our system helps taxi companies regularly analyze data for better decision-making. The usage scenarios show several abnormal areas with high taxi demand or poor travel experience, which can provide insights to analysts and guide the scheduling of taxis. For example, it can coordinate the relationship between drivers and passengers and improve the income of drivers by reasonable incentives to guide drivers' cruises. The company needs to improve the environment of taxi riding, i.e., setting up riding areas or pick-up points at appropriate locations, especially in places with dense traffic such as scenic spots and stations.
In terms of drivers, there are several suggestions for them to cruise. First, plan cruise routes according to peak times in dense traffic areas. It helps prevent missing crowds and avoid congestion on the way. Second, focus on the surrounding POI types. For example, passengers in some areas prefer ``Living" and ``Entertainment" POI. }

\textbf{Partition of the urban area.} 
As mentioned from E1 and E2, in {\name} we partition the urban area equally into hexagons regardless of the contextual information. This is enough for the current method to some extent. However, the current segmentation methods may not be effective, if we want to integrate more mainstream deep learning methods in the future. Furthermore, the previous literature~\cite{zeng2020revisiting} mentioned that traffic aggregations might depend on the shapes and scales of the spatial partition units, for example, the MAUP~\cite{openshaw1984modifiable}~\cite{gehlke1934certain}. Therefore, further exploration between the partition of the area and the result for recommendation is still our future task.

\textbf{Analysis of temporal pattern.} \gsx{E3 pointed out that in the multiscale temporal chart, the ticks at the same time cannot be aligned as the rectangle length represents both flow and time. This puts a burden on the analysis of temporal patterns. Figuring out how to align the ticks will be our future work.}

\textbf{Scalability.}
{\name} is designed for analyzing the pick-up point selection behavior of passengers. With a large number of taxi GPS data, it is easy to cause scalability issues. For better in-depth exploration, we have adopted a three-scale exploration strategy, namely city, region, and point scales. While the map design and the beeswarm graph will show visual clutters when we explore the information of a large area. We have provided another option, the stacked bar chart, to ease visual clutters. To better handle the scalability issues, we plan to explore more designs and adopt some automatic methods to filter unnecessary information.

\textbf{Generality.}
Although we focus on analyzing taxi data, our analytical pipeline and strategy can be easily adopted to other similar spatial-temporal data, such as bus data, and telco data. The design for comparison can also be applied to other applications for comparing temporal data.

\textbf{Evaluation.} 
Our system, {\name}, is currently evaluated with only three expert users. To better evaluate the usability and effectiveness of our system, a long-term study with more domain experts are needed, which is left for future work.

\section{Conclusion}
\label{sec:conclusion}

In this paper, we propose {\name}, an interactive visual analytics system that helps users visually explore passengers' behavior of pick-up point selection. 
Several well-designed visualizations and interaction techniques are combined to facilitate multi-scale exploration at city, region and point scales. Coordinated contrast views are provided to compare different patterns in different regions. We propose a set of criteria for examining the performance of each pick-up point. In the end, we demonstrate the effectiveness of our system with three case studies on a real-world taxi dataset in Shenzhen and interviews with three domain experts. The results show that our system is useful to explore the pick-up points selection behavior of passengers and provide guidance for empty taxi cruising. In the future, we will integrate recommendation algorithms for better suggestions for users.

\acknowledgments{
We would like to thank our domain experts and the anonymous reviewers for their insightful comments. This work is supported by the 100 Talents Program of Sun Yat-sen University.}

\bibliographystyle{abbrv-doi}

\balance
\bibliography{template}
\end{document}